\documentclass[french,arxiv,superscriptaddress,amsmath,amssymb,longbibliography,onecolumn,floatfix,11pt]{revtex4-2}   
\usepackage{amsmath}
\usepackage{graphicx}  
\usepackage{dcolumn}   
\usepackage{bm}        
\usepackage{amssymb}   
\usepackage[normalem]{ulem}

\usepackage{tikz}
\usetikzlibrary{arrows, arrows.meta}
\usepackage{hyperref}
\hypersetup{
    colorlinks=true,
    linkcolor=blue,
    filecolor=blue,      
    urlcolor=black,
    citecolor=blue,
        }
\usepackage{subfigure}
\begin{document}
\title{Red blood cells aggregates transport for finite concentration}
\author{Mehdi Abbasi}
\email{mehdi1abbasi@gmail.com}
\affiliation{Aix Marseille Université, CNRS, Centre Interdisciplinaire de Nanoscience de Marseille, Turing Centre for Living Systems, 13009 Marseille, France}
\affiliation{Université Grenoble Alpes, CNRS, LIPhy, F-38000 Grenoble, France}
\author{Chaouqi Misbah}
\email{chaouqi.misbah@univ-grenoble-alpes.fr}
\affiliation{Université Grenoble Alpes, CNRS, LIPhy, F-38000 Grenoble, France}

\begin{abstract}
Red blood cells (RBCs) are responsible for transporting oxygen and various metabolites to tissues and organs, as well as removing waste. Several cardiovascular diseases can impair these functions. For instance, in diabetes, increased RBC aggregation can lead to blood occlusion, thereby depriving tissues of efficient oxygen delivery. Interestingly, RBC adhesion occurs not only in disease states but also under physiological conditions, with the key difference being that adhesion is reversible in healthy situations.
This paper focuses on numerical simulations in 2D, exploring different adhesion energies (both physiological and pathological) alongside varying flow strengths and hematocrit levels. A systematic analysis of RBC flux and viscosity is conducted. A remarkable finding is that moderate adhesion energy (within the physiological range) enhances RBC transport, thereby improving oxygen delivery to tissues. This provides insight into why RBC adhesion is present under normal conditions. Conversely, increasing adhesion energy beyond a certain point causes a collapse in RBC flux, thus reducing oxygen transport.
We provide a basic explanation for the non-monotonic effect of adhesion energy on blood flow efficiency. This finding serves as an initial step in understanding the impact (both positive and negative) of RBC adhesion before addressing this issue in complex networks inspired by realistic microvasculatures.
\end{abstract}
\maketitle

\section{Introduction}
Transport of red blood cells (RBCs) is crucial in hemodynamics, as RBCs carry oxygen from the lungs to the tissues and remove carbon dioxide from the tissues to the lungs. 

Several  diseases, such as diabetes, hypercholesterolemia, and drepanocytosis are often associated with alterations in blood properties\cite{bateman2003bench}. For example, drepanocytosis \cite{di2016dense, lemonne2012increased}, characterized by a stiffened RBC cytoplasm under low oxygen conditions, impairs proper RBC flow in small blood vessels due to the formation of occlusions. One prominent feature in blood flow is the susceptibility of RBCs to form aggregates, due to  adhesion force provoked by plasma proteins, such as fibrinogen\cite{brust2014plasma}. In physiological conditions, the human fibrinogen level typically ranges from approximately 1.8 to 4 mg/ml \cite{brust2014plasma}. Under physiological conditions, the RBCS aggregation is reversible; RBCs aggregate and dis-aggregate in response to blood flow stress, thus maintaining blood flux and perfusion to tissues and organs. 
However, under pathological conditions  ( in cardiovascular diseases, such as diabetes), the aggregation may become irreversible, impairing blood  perfusion  to tissues and organs\cite{comeglio1996blood}. Understanding the transport of RBCs and their behavior under different flow conditions, particularly under pathological conditions, has the potential to guide   the understanding of far-reaching consequences of RBCs aggregates.

Two main mechanisms are proposed to explain the  RBC-RBC adhesion. i- The bridging mode, which is l is based on the assumption that the proteins responsible for aggregation (e.g. fibrinogen) adsorb onto the RBC surface to form a cross-link with the nearby RBC via receptor sites \cite{chien1973ultrastructural, brooks1988mechanism}. ii- the depletion model based on the fact that the molecules (due to their thermal fluctuations) have a depletion zone adjacent to  the RBCs surface. When the surfaces of two RBCs are close enough to each other (a distance of the order of the depletion zone), the region separating the two surface are  depleted in aggregating molecules, creating thus an   osmotic  attractive force between adjacent RBCs \cite{baskurt2011red}. Note that neither of these two mechanisms has received universal acceptance and both mechanisms may be involved in this phenomenon to different degrees depending on the nature of the aggregating molecules and the receptor sites on the RBC membrane. The negative surface charge of RBCs leads to a surface potential which in turn creates an electrostatic repulsive force, and together with the attractive part, yields  to an equilibrium distance between cells.

The understanding of RBCs aggregation in static conditions and  under flow has received a significant attention both experimentally and by means of numerical simulation \cite{abbasi2021erythrocyte, baskurt2011red, chien1973ultrastructural, hoore2018effect, flormann2017buckling, brust2014plasma, claveria2016clusters}. From the simulation side, the situation turned out to be subtle, even in the simplest configuration of aggregation of a pair of cells (doublet). Indeed, it has been found  that when a single cell performs tumbling  the doublet formed due to adhesion (even very weak) remains stable even under a
very strong shear rate\cite{abbasi2021erythrocyte}. The understanding of blood flow properties at finite hematocrit in the presence of adhesion among RBCs remains a largely unexplored research area. This constitutes the main objective of the present study.

The flux of rigid or soft particles (droplets, vesicles, capsules...) is a complex function of the particle concentration, depending strongly on the suspension structure induced by interactions with walls and among particles. The optimality in blood flow is an interesting problem in hemorheology. Hematocrite is likely to be optimized due to its effect on nutriments transport. The problem of optimal hematocrit has been the subject of many studies \cite{lipowsky1980vivo, barbee1971prediction, birchard1997optimal, hedrick1986blood, linderkamp1992blood}. Recently, it has been shown that the RBCs transport depends strongly on the RBCs properties, the vessel diameter and the flow strength \cite{farutin2018optimal}. Gou et al \cite{gou2021red} have shown that the quantity of the ATP released by RBCs suspensions depends strongly on the hematocrit, the ATP released per cell increases to a maximum which corresponds to an optimal physiological hematocrit. Optimality in living systems is a delicate question that must be approached with care. In \cite{farutin2018optimal, gou2021red}, the authors found that the flux of RBCs (a model of RBCs) and ATP release by vesicles both show an optimal value with the RBCs concentration. 

Here we will study by numerical simulations the behavior of RBCs suspensions in a straight channel, by exploring several hematocrits and adhesion energies in both physiological and pathological ranges. We will analyze the RBCs flux and viscosity as functions of the adhesion energy,  the flow strength and hematocrit. A remarkable feature reported here is that the RBCs flux exhibits a non monotonic behavior. A low enough adhesion energy (corresponding to physiological values) it is found that the RBCs flux is boosted. For a given flow strength, we find an increase of RBCs flux up to about 20$\%$. This means that a moderate adhesion among RBCs is not detrimental, but rather favors oxygen delivery  to tissues and organs. For a larger adhesion energy the RBCs flux exhibits a decline.
We will provide a basic  explanation for  the non-monotonic behavior of the RBCs flux with the adhesion energy.

\section{Model and simulation method}
This study was conducted using a two-dimensional vesicle model. Interestingly, certain shapes and dynamics previously attributed to the cytoskeleton \cite{mauer2018flow, lanotte2016red, fischer1978tank, minetti2019dynamics, Fischer2013, dupire2012full, abkarian2007swinging, tomaiuolo2009red} were successfully reproduced in 2D \cite{abbasi2022dynamics, kaoui2009red, kaoui2011complexity, biben2011three, aouane2014vesicle, Kaoui2009VesiclesUS, agarwal2020stable, trozzo2015axisymmetric}, thereby challenging the hypothesis that the cytoskeleton is essential for their emergence.

\subsection{Lattice Boltzmann method}
The Navier-Stokes equations are solved in the limit of small Knudsen and Mach numbers using the lattice Boltzmann method (LBM) which is based on the discretisation of the Boltzmann-BGK \cite{succi2001lattice, aidun2010lattice, bhatnagar1954model, kruger2017lattice} equation in time and space. In LBM, the fluid is seen as cluster of pseudo-fluid particles. The main quantity in LBM is the distribution function $f_{i}(\mathbf{x},t)$, which gives the probability of finding a particle at position $\mathbf{x}$ and time $t$ having a microscopic velocity $\mathbf{c}_{i} = (c_{ix}, c_{iy})$, where $i = 1,...,Q$, on a regular D-dimensional lattice in discrete time steps $\Delta t$. We consider here a $D2Q9$ model corresponding to a two dimensional lattice with $Q = 9$ velocities. The lattice Boltzmann equation has the following form:
\begin{equation}
f_{i}(\mathbf{x} + \mathbf{c}_{i} \Delta t) - f_{i}(\mathbf{x},t) = \Omega_{i}(\mathbf{x},t) + F_{i}(\mathbf{x},t)\Delta t,
\end{equation}
Where $F_{i}(\mathbf{x},t)$ is a bulk force term to be specified below and $\Omega_{i}(\mathbf{x},t)$ is the collision operator which is written as follows:
\begin{equation}
\Omega_{i}(\mathbf{x},t) = - \frac{\Delta t}{\tau} [f_{i}(\mathbf{x},t)-f_{i}^{eq}(\mathbf{x},t)],
\end{equation}
$f_{i}^{eq}(\mathbf{x},t)$ is the equilibrium distribution, is obtained from an approximation of the Maxwell distribution and can be expressed as:
\begin{equation}
f_{i}^{eq}(\mathbf{x},t) = \omega_{i} \rho [1 + \frac{(\mathbf{c_{i}} \cdot \mathbf{u})}{c_{s}^{2}} + \frac{1}{2} \frac{(\mathbf{c_{i}} \cdot \mathbf{u})^{2}}{c_{s}^{4}} - \frac{1}{2} \frac{|\mathbf{u}|^{2}}{c_{s}^{2}}],
\end{equation}
where $\tau$ is a dimensionless relation time, related to the kinematic viscosity of the fluid $\nu = \Delta t c_{s}^{2} (\tau - 1 / 2)$. Here $c_{s} = (1/\sqrt{3})(\Delta x / \Delta t)$ called the lattice speed of sound in D2Q9 model, $\omega_{i}$ are lattice weights, $\Delta x$ is the lattice constant, $\rho$ the fluid density, $\mathbf{u}$ is the velocity fields as a truncated expansion of the Maxwell–Boltzmann distribution (valid at small Mach number, $Ma = |\mathbf{u}| / c_{s} \ll 0.1$). For the $D2Q9$ LBM model, the nine lattice velocities $\mathbf{c}_{i}$ and weight factors $\omega_{i}$ are $\mathbf{c}_{i} = (0,0)$, $\omega_{i}=4/9$ for $i=0$; $\mathbf{c}_{i} = (cos[i-1] \pi /2, sin[i-1] \pi /2)$, $\omega_{i} = 1/9$ for $i = 1, 2, 3, 4$; and $\mathbf{c}_{i} = (cos[2i - 9] \pi /2, sin[2i-9] \pi/2)$, $\omega_{i} = 1/36$ for $i= 5, 6, 7, 8$. The viscosity $\eta$ of the fluid can be varied in the LBM by adopting two different relation time inside and outside the vesicles in the BGK expression (see \cite{kaoui2016two} for the implementation of the viscosity contrast). In our simulation, we set $\tau = 1$, $\Delta t = 1$ and $\Delta x = 1$, yielding $\nu = 1/6$ for the viscosity in the LBM units. Then a viscosity contrast is created thanks to a different choice of $\tau$ for the fluid inside the vesicle. For example, we set $\tau = 5.5$ for the fluid inside the vesicle to have viscosity contrast $\lambda = 10$.

Once the distribution function $f_{i}$ is determined from eq-, it becomes possible to extract the macroscopic quantities, the fluid density $\rho$  can be computed as :

\begin{equation}
\rho = \sum_{i} f_{i},
\end{equation}

The fluid velocity is expressed as :

\begin{equation}
\mathbf{u} \rho = \sum_{i} f_{i} \mathbf{c_{i}} + \frac{1}{2} \Delta t \mathbf{f},
\end{equation}

\subsection{Membrane model}

In the present article, we consider a set of 2D phospholipid vesicles as a model of RBCs. 2D Vesicles model has demonstrated the ability to capture several various characteristics shown by RBCs. Notably, shapes like slipper and parachute \cite{guckenberger2018numerical, kaoui2009red}, dynamics like tumbling and tank treading \cite{biben2011three, kaoui2009vesicles}, multilobe shapes \cite{abbasi2022dynamics, lanotte2016red, mauer2018flow}, ATP released by RBCs \cite{gou2021red}, dynamic of RBCs doublet aggregates \cite{abbasi2021erythrocyte}, rheology of confined suspension \cite{thiebaud2014prediction} and optimal cell transport \cite{farutin2018optimal}. We consider a suspension of vesicles inside a straight channel, bounded by two rigid walls located at $y=0$ and $y=W$, where $W$ is the channe width as represented in Fig \ref{fig5:0}, the suspension is subject to a parabolic flow Eq.\ref{poiss}. 

The force $\mathbf{f}$ acting from vesicle membranes on the surrounding fluid can be obtained from the derivative of the following energy, which is the sum of four terms : 

\begin{itemize}
\item The stretching energy \cite{tsubota2006particle, tsubota2010effect} : 

\begin{equation}
E_{l} = \frac{1}{2} \mathit{k_{l}} \sum_{i = 1}^{N} \left(\frac{l_{i}- l_{0}}{l_{0}} \right)^{2}
\end{equation}
where $\mathit{k_{l}}$ is a springlike constant, $l_{i}$ and $l_{0}$ are the actual and the equilibrium spring length respectively.

\item To maintain the area $s$ of the vesicle internal fluid at a constant value $s_{}$, we adopt an additional penalization term of area in the potential energy \cite{tsubota2006particle, tsubota2010effect} : 
\begin{equation}
E_{s} = \frac{1}{2}\mathit{k_{s}}\left( \frac{s - s_{0}}{s_{0}}\right) 
\end{equation}
where $\mathit{k_{s}}$ is the a constant chosen as large as possible in order to ensure a constant enclosed area

\item The bending energy is written as \cite{tsubota2006particle, tsubota2010effect} : 
\begin{equation}
E_{b} = \frac{1}{2} \mathit{k_{b}} \sum_{i=1}^{N} tan^{2}\left(\frac{\theta_{i}}{2} \right) 
\end{equation}
where $\mathit{k_{b}}$ is the bending constant and $\theta_{i}$ is the supplementary angle between two springs connected to the membrane point $i$.

\item The adhesion energy based on the Lennard-Jones potential : 
\begin{equation}
E_{adh} = \epsilon\left( -2\left( \frac{\sigma}{r} \right)^{6} + \left( \frac{\sigma}{r} \right)^{12} \right)
\end{equation}
where $r$ is the separation between two nodes belonging to two adhered vesicles,  $\epsilon$ is the minimum energy corresponds to the equilibrium distance $\sigma$.
\end{itemize}

The membrane force on each point at each time step can computed as follows : 
\begin{equation}
\mathbf{f} = - \left( \frac{\partial(E_{b} + E_{l} + E_{s} + E_{adh})}{\partial \mathbf{r}}   \right)
\end{equation}

The force is coupled to LBM solver of the fluid flow through the use of IB method via the following equation  \cite{kaoui2011two, zhang2007immersed, feng2004immersed, shen2017interaction}: 

\begin{equation}
\mathbf{F}(\mathbf{r}_{f}) = \sum_{m} \mathbf{f}(\mathbf{r}_{m}) \delta (\mathbf{r}_{f} - \mathbf{r}_{m})
\end{equation}
where $\delta(\mathbf{r})$ is a regularized Dirac delta function. The membrane velocity is calculated by adopting a no-slip boundary condition. Similarly the velocity is written as follow : 
\begin{equation}
\mathbf{u}(\mathbf{r}_{m}) = \sum_{f} \mathbf{u}(\mathbf{r}_{f}) \delta (\mathbf{r}_{m} - \mathbf{r}_{f})
\end{equation}

\subsection{Simulation strategy}

Initially, the cells are distributed randomly inside the channel, and then a parabolic flow (Poisseuille flow) Eq.\ref{poiss} is applied to the suspension as shown in Fig \ref{fig5:0}.

\begin{equation}
\left\{
    \begin{array}{ll}
        u_{x}^{0} = 4 u_{m} y ( W - y ) / W^{2} \\
        u_{y}^{0} = 0
    \end{array}
\right\}
\label{poiss}
\end{equation}
where $u_{m}$ is the maximal velocity, $W$ is the channel width.

\begin{figure}[hbtp]
\centering
\includegraphics[scale=0.65]{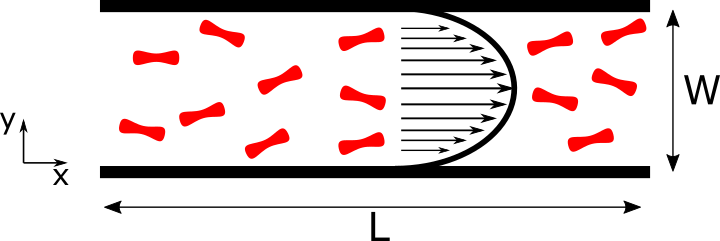}
\caption{\label{fig5:0} Schematic of the simulation.}
\end{figure}

The quantities of interest are averaged over time once the simulation reaches the steady state. The RBCs flux is represented by means of the cell flow rate $Q_{c}$ normalized by the flow rate of the cell-free fluid $Q_{0}$ under the same flow strength. The cell flow rate is measured by counting the number of cells passing throw a section in the channel during a fixed time interval. 

The effective viscosity is defined as: 
\begin{equation}
\eta_{eff} P_{c} = \eta_{0} P_{0} 
\end{equation} 
Where $P_{c}$ and $P_{0}$ are the average flux of the suspension (not to be confused by cells flux, $Q_c$) and the flux of the pure fluid  (with viscosity $\eta_{0}$) respectively. 

The normalized viscosity can be expressed as follow:
\begin{equation}
[\eta] = \frac{\eta_{eff} - \eta_{0}}{\eta_{0} \phi}
\label{poiseff}
\end{equation}
where $\phi_{t} = n \pi R_{0}^{2} / (LW)$ represents concentration of the cell suspension, $n$ the number of the cells in the suspension.

All the quantities calculated in this article are averaged over time in the steady state.

\subsection{Dimensional numbers}

Dimensionless numbers are used to describe the vesicle and the flow characteristics:

\begin{itemize}
\item The capillary number: allows to quantify the flow strength over bending rigidity of the membrane
\begin{equation}
C_{a} = \dfrac{\eta_{0} <\dot{\gamma}> R_{0}^{3}}{\mathit{k}} \equiv \dot{\gamma} \tau_{c}
\end{equation}
$<\dot{\gamma}> = 2u_{m}/W$ is the averaged shear rate along the $y$ direction and $k$ is the bending rigidity modulus. 
\item The viscosity contrast: the ratio between the viscosities of the internal and external fluids
\begin{equation}
\lambda = \dfrac{\eta_{1}}{\eta_{0}}
\end{equation}
\item The reduced area: combining the vesicle perimeter  $L$  and its enclosed area $A$ 
\begin{equation}
\tau = \dfrac{(A/ \pi)}{(L/2 \pi)^{2}}
\end{equation}
\item The dimensionless adhesion energy : 
\begin{equation}
\bar{\varepsilon}_{adh} =\dfrac{{\varepsilon_{adh}} R_{0}^{2}} {\mathit{k}}
\end{equation} 
Where $\varepsilon_{adh} \simeq 1.6862 h \varepsilon$, is the interaction energy of adhesion between two infinite plates\cite{abbasi2021erythrocyte}.
\end{itemize}

\begin{table}[hbtp]
\begin{tabular}{ l c c c c c}
Fibrinogen level $mg/ml$ | & {\color{blue}0.89} & {\color{blue}2.39} & {\color{orange}4.167} & {\color{red}8.09} \\
Adhesion energy $\mu J/ m^{2}$ | & {\color{blue}-1.88} & {\color{blue}-3.748} & {\color{orange}-4.92} & {\color{red}-6.56} \\
$\bar{\varepsilon}_{adh}$ | & {\color{blue}42.3} & {\color{blue}61.1} & {\color{orange}84.33} & {\color{red}147.7} \\
\end{tabular}
\caption{\label{tab:1} 
Fibrinogen level versus Interaction energy between two RBC measured using atomic force microscopy \cite{brust2014plasma}. Blue, orange, and red represent the physiological case, intermediate regime, and pathological case, respectively.}
\end{table}

Throughout this article, the reduced area value will be set to $0.65$ (inspired by that of human RBCs). 
\section{Results and discussion}

In previous study \cite{abbasi2021erythrocyte}, it has been shown that at high enough viscosity contrast $(\lambda = 10.0)$ the erythrocyte-erythrocyte doublet becomes quite stable even for high enough applied shear stress. In this part we will discuss some features of  the aggregation of RBCs for different regimes of viscosity contrast. Then our main goal is to evaluate the effect of aggregation on the RBCs flux and the suspension viscosity.

\subsection{Reversible and irreversible aggregation of RBCs}
\label{rev_irrev}

\begin{figure}[hbtp]
\centering
\begin{subfigure}
\centering
\includegraphics[scale=0.25]{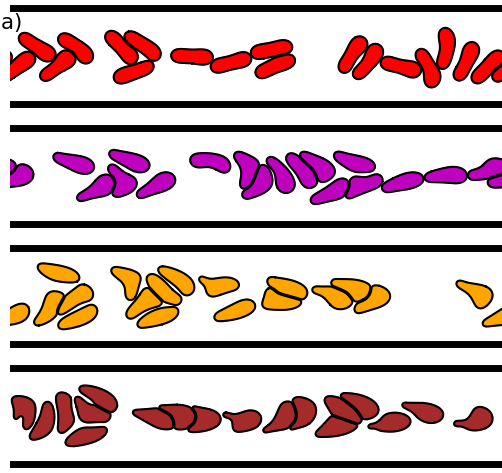}
\end{subfigure}
\hspace{0.1cm}
\begin{subfigure}
\centering
\includegraphics[scale=0.25]{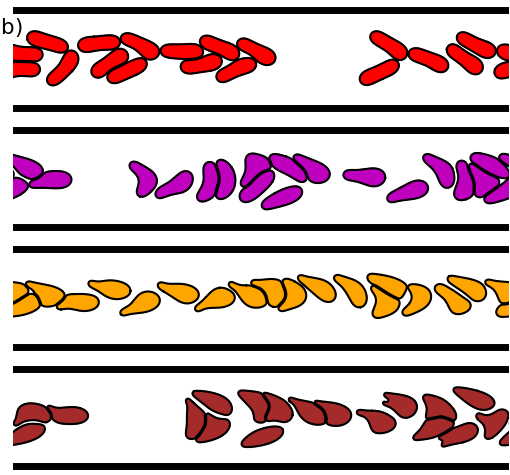}
\end{subfigure}
\hspace{0.1cm}
\begin{subfigure}
\centering
\includegraphics[scale=0.25]{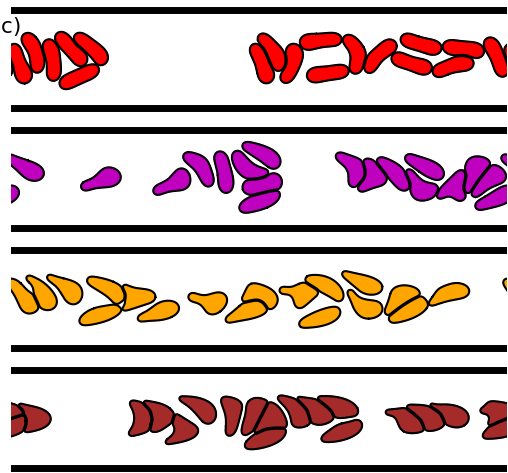}
\end{subfigure}
\caption{\label{fig5:-2} Snapshots showing the spatial configurations of the cells at different time steps. a) $1500 \tau_{c}$, b) $1750 \tau_{c}$ and c) $2000 \tau_{c}$. Here the capillary number is set to $C_{a} = 25.0$ , the cells concentration is $\phi_{t} = 0.2$ and the viscosity contrast is set to $\lambda = 1.0$. The colors show different macroscopic dimensionless adhesion energy values. Red : $36.81$, purple : $73.63$, yellow : $147.27$ and brown : $220.91$}
\end{figure}
\begin{figure}[hbtp]
\centering
\begin{subfigure}
\centering
\includegraphics[scale=0.25]{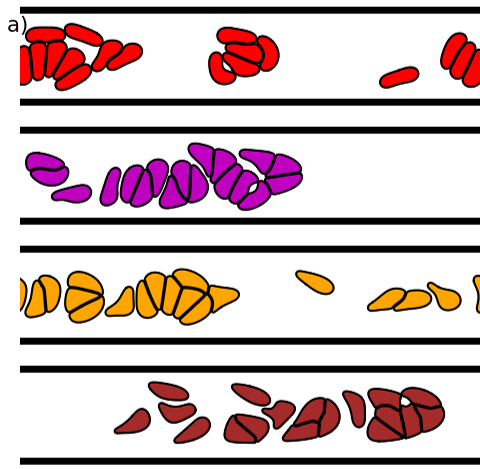}
\end{subfigure}
\hspace{0.1cm}
\begin{subfigure}
\centering
\includegraphics[scale=0.25]{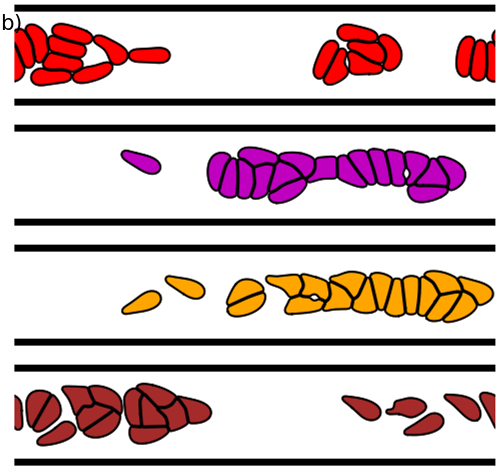}
\end{subfigure}
\hspace{0.1cm}
\begin{subfigure}
\centering
\includegraphics[scale=0.25]{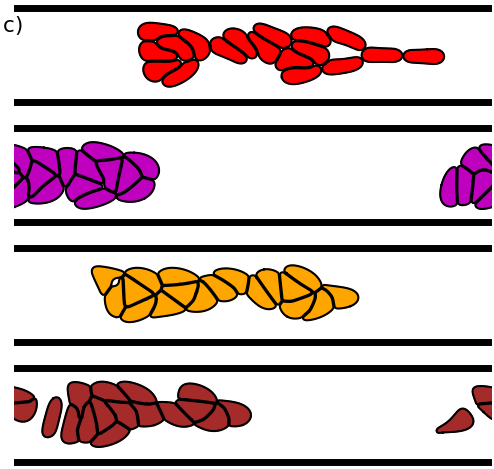}
\end{subfigure}
\caption{\label{fig5:-1} Snapshots showing the spatial configurations of the cells at different time steps. a) $1500 \tau_{c}$, b) $1750 \tau_{c}$ and c) $2000 \tau_{c}$. Here the capillary number is set to $C_{a} = 25.0$, the cells concentration is $\phi_{t} = 0.2$ and the viscosity contrast is set to $\lambda = 10.0$. The colors shows different macroscopic adhesion energy values. Red : $36.81$, purple : $73.63$, yellow : $147.27$ and brown : $220.91$. The last panel corresponds to a pathological adhesion energy; see table \ref{tab:1}.}
\end{figure} 

Under  physiological conditions, the aggregation process is believed to be reversible. In contrast, under  pathological conditions the aggregation may become  irreversible. In our recent study\cite{abbasi2021erythrocyte}, we found that the RBCs doublet aggregates are stable even for physiological condition of adhesion energy and even for high enough applied shear stress. The non dissociation is due to the high viscosity contrast of RBCs \cite{abbasi2021erythrocyte}, as expected to occur for some diseases (e.g. malaria). Here we show that the non dissociation of cells also occurs for RBCs suspension under Poiseuille flow. In figure \ref{fig5:-2}, we see that for low enough viscosity contrast the cells form aggregates in a reversible way as long as the adhesion energy remains within the physiological range: they continuously adhere to each other and they dissociate during time. The snapshots of reversible aggregation are shown in figure \ref{fig5:-2} with red and purple color. When the adhesion energy is high enough (above physiological values) the aggregation becomes irreversible (see figure \ref{fig5:-2}, yellow and brown color). The relation between dimensionless adhesion energy and fibrinogen concentration is given in table \ref{tab:1}, showing both physiological and pathological adhesion.

\begin{figure}[hbtp]
\centering
\includegraphics[scale=0.5]{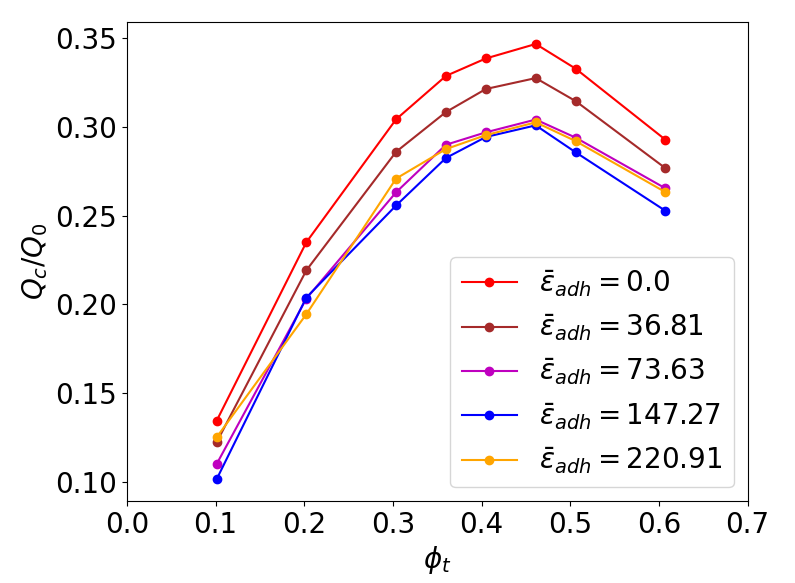}
\caption{\label{fig5:0} Normalized cell flow rate as a function of volume fraction of cell for different dimensionless macroscopic adhesion energy. The simulation data are shown as dots. Here the viscosity contrast is $\lambda = 10.0$ and the capillary number is $C_{a} = 25.0$}
\end{figure} 

In contrast with the case of low viscosity contrast discussed above, for high viscosity contrast, we observe that once the aggregates form they remain quite stable; they may never  dissociate, as shown in figure \ref{fig5:-1}, even for low adhesion energy values (within physiological range). This is due to the robustness of RBC aggregation, due to absence of tank-treading,  as analyzed in detail in \cite{abbasi2021erythrocyte}.

In Fig \ref{fig5:0}, we show the normalized cell flow rate as function of volume fraction of cell for different dimensionless adhesion energy. We clearly see that the irreversibility of aggregation due to high value of viscosity contrast ($\lambda = 10.0$) has a dramatic effect on RBCs transport. As we will explain in the next subsection \ref{transport}, in the absence of adhesion the cell transport increases to a maximum and then decreases. The formation of robust aggregates affects strongly the RBC transport.

We will first focus on the viscosity contrast value  $\lambda = 1.0$, which is the regime where reversible aggregation occurs (if adhesion energy is not too high) and we will investigate the effect of the adhesion energy on the RBCs transport and RBCs suspension viscosity. We will briefly discuss the case of large viscosity contrast $\lambda=10$, where aggregation becomes irreversible.

\subsection{The effect of adhesion energy on the normalized cell flux}
\label{sec1}
\begin{figure}[hbtp]
\centering
\begin{subfigure}
\centering
\includegraphics[scale=0.25]{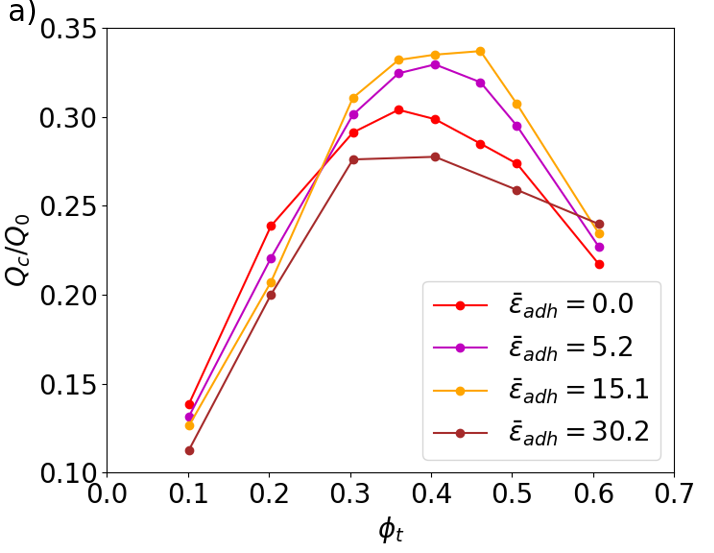}
\end{subfigure}
\hspace{0.1cm}
\begin{subfigure}
\centering
\includegraphics[scale=0.25]{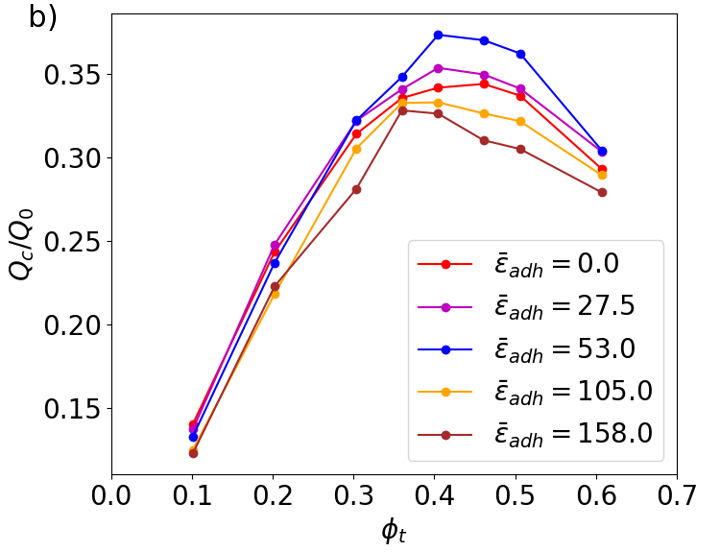}
\end{subfigure}

\begin{subfigure}
\centering
\includegraphics[scale=0.25]{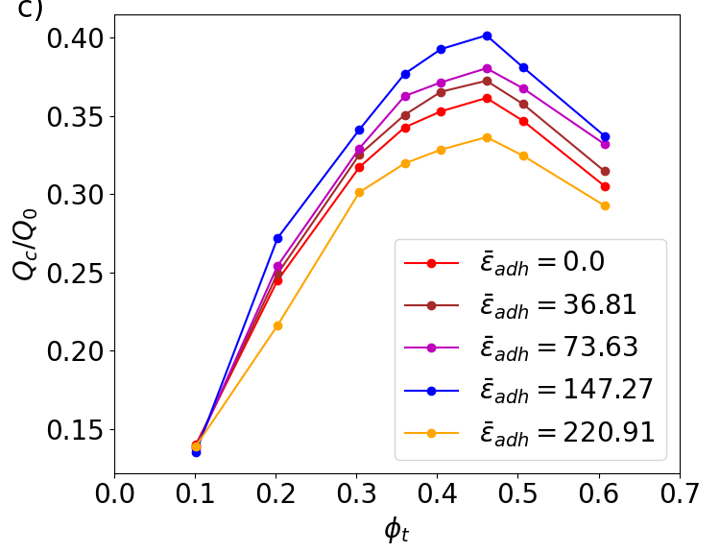}
\end{subfigure}
\hspace{0.1cm}
\begin{subfigure}
\centering
\includegraphics[scale=0.25]{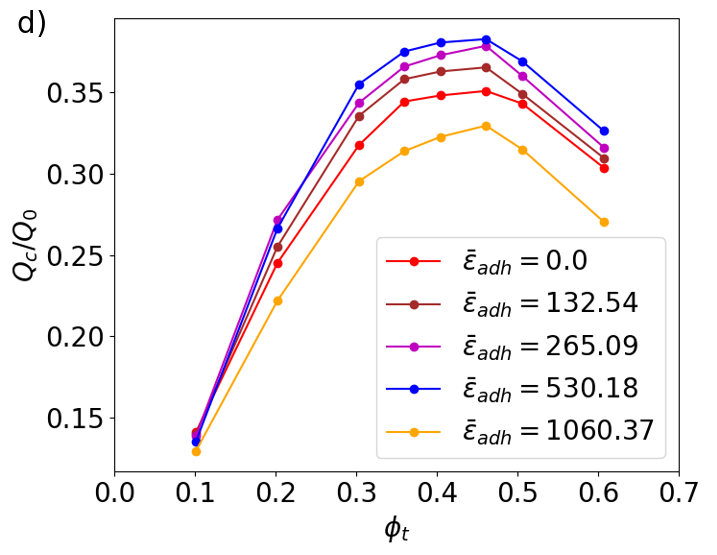}
\end{subfigure}
\caption{\label{fig5:1} Normalized cell flow rate as a function of volume fraction for channels with different dimensionless macroscopic adhesion energy. The simulation data are shown as dots. (a) $C_{a} = 0.9$, (b) $C_{a} = 9.0$, (c) $C_{a} = 25.0$, (d) $C_{a} = 90.0$.}
\end{figure}

Here, we investigate the effect of the adhesion energy on the RBC flux for different capillary numbers. Figure \ref{fig5:1} summarizes  the results. In the absence of adhesion, where $\bar{\epsilon}_{adh} = 0.0$ (the relation between dimensionless adhesion energy and fibrinogen concentration is given in table \ref{tab:1}), our results are in a good quantitative agreement with simulation results of Farutin et al. \cite{farutin2018optimal}. The normalized cell flux $Q_{c} / Q_{0}$ exhibits a maximum as a function of the cell concentration for different capillary numbers. It is reported \cite{farutin2018optimal} that the optimal hematocrit is sensitive to the capillary number, vessel diameter, reduced area. The optimal hematocrit values found\cite{farutin2018optimal} for vessel sizes corresponding to macrocirculation and intermediate microcirculation (arterioles) are close enough to the corresponding physiologically values.

\begin{figure}[hbtp]
\centering
\includegraphics[scale=0.5]{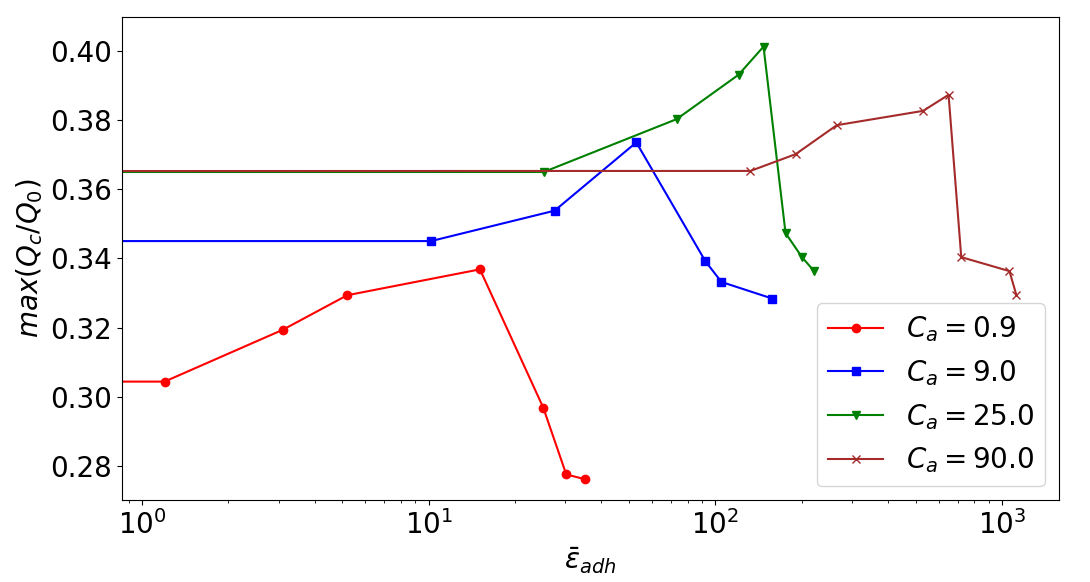}
\caption{\label{fig5:2} The maximal value of the normalized flow rate as function of the dimensionless macroscopic adhesion energy for different values of capillary number. The horizontal axis is in logarithmic scale.}
\end{figure}

Now, we focus on the effect of adhesion between cells (formation of aggregates) on the cell flux and optimal hematocrit. The channel width value is fixed at $22.5 \mu m$ (a typical value in microcirculation). The simulations are performed for different values of capillary numbers. In Fig. \ref{fig5:1}, we see that including adhesion in our model does not affect qualitatively the shape of the flux: the existence of an optimal hematocrit for RBC transport is robust even in the presence of adhesion between RBCs. At very low capillary number $C_{a}$ (where cell deformation is weak) the optimal hematocrit value is shifted towards high concentration values when the adhesion energy $\bar{\epsilon}_{adh}$ is increased. Surprisingly, we found that the maximal value of the normalized cell flux $Q_{c} / Q_{0}$ is not monotonic with the adhesion energy $\bar{\epsilon}_{adh}$. A low enough adhesion boosts RBCs flux. When adhesion energy is large enough the flux attains a maximum; the corresponding adhesion energy is referred to "optimal energy". Beyond this value the maximum RBCs flux declines.
 The optimal adhesion energy value depends strongly on the capillary number as shown in Fig. \ref{fig5:2}.   
 Under physiological conditions it is known that RBCs aggregates form and dissociate, as found here. The interesting fact obtained here is that adhesion (leading to reversible aggregates), if not too large, enhances the RBCs flux, meaning that this type of aggregates formation would be  beneficial for perfusion, a quite counter-intuitive result (see explanation below). Note that the increase in maximum flux may be quite large, as shown in  Fig. \ref{fig5:2}. For a zero adhesion, the smallest relative flux is about 0.3, and can attain, in the presence of adhesion, values up to  0.4 (for $\bar{\epsilon}_{adh}\sim 200$ and $Ca=25$). This corresponds to an increase of the maximum RBCs flux by $30 \%$, which is quite significant. In Figs. \ref{fig5:1} and \ref{fig5:2}, we observe that for a high capillary number ($C_{a} = 90.0$), the adhesion energy values explored are higher compared to the other cases of capillary number ($C_{a} = 25.0, 9.0, 0.9$). This is attributed to the competition between aggregation and disaggregation forces: as the disaggregation forces increase, higher aggregation forces are required to form aggregates. Consequently, the values explored in Figs. \ref{fig5:1} and \ref{fig5:2} (blue and yellow curves) represent an extreme condition.


\begin{figure}[hbtp]
\centering
\begin{subfigure}
\centering
\includegraphics[scale=0.45]{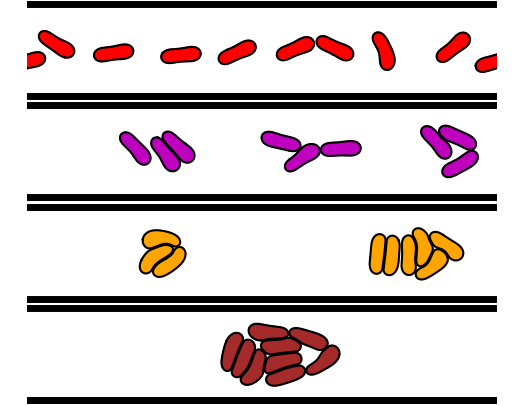}
\end{subfigure}
\hspace{0.1cm}
\begin{subfigure}
\centering
\includegraphics[scale=0.45]{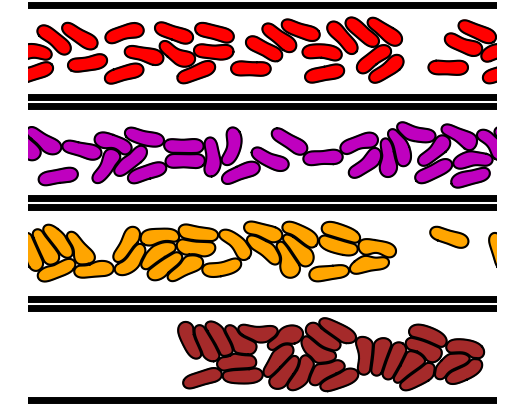}
\end{subfigure}

\caption{\label{fig5:3} Snapshots showing the spatial configurations of the cells for different macroscopic adhesion energy values. The colors show different macroscopic adhesion energy values. Red : $0.0$, purple : $5.2$, yellow : $15.1$ and brown : $220.91$. Here the capillary number is set to $C_{a} = 0.9$ and the viscosity contrast is set to $\lambda = 1.0$. Left : $\phi_{t} = 0.1$. Right : $\phi = 0.4$}
\end{figure}

\begin{figure}[hbtp]
\centering
\begin{subfigure}
\centering
\includegraphics[scale=0.5]{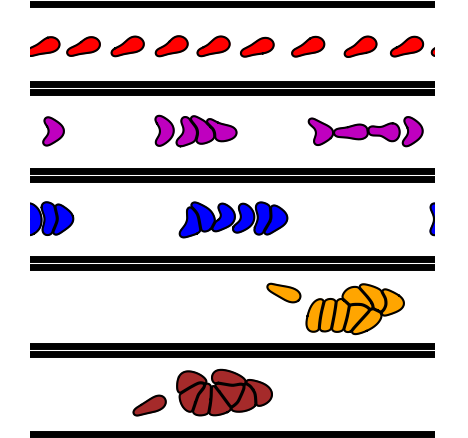}
\end{subfigure}
\hspace{0.1cm}
\begin{subfigure}
\centering
\includegraphics[scale=0.5]{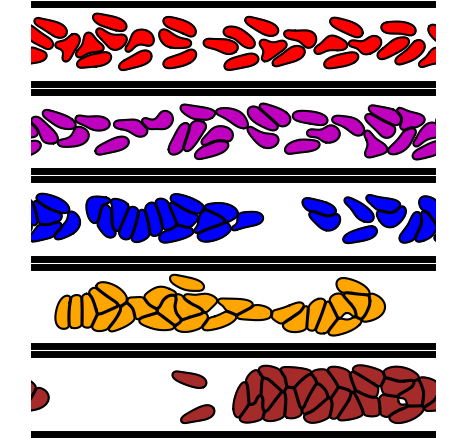}
\end{subfigure}

\caption{\label{fig5:4}  Snapshots showing the spatial configurations of the cells for different macroscopic adhesion energy values. The colors show different macroscopic adhesion energy values. Red : $0.0$, purple : $27.5$, blue : $53.0$ ,yellow : $105.0$ and brown : $158.0$. Here the capillary number is set to $C_{a} = 9.0$ and the viscosity contrast is set to $\lambda = 1.0$. Left : $\phi = 0.1$. Right : $\phi = 0.4$}
\end{figure}

\begin{figure}[hbtp]
\centering
\includegraphics[scale=0.85]{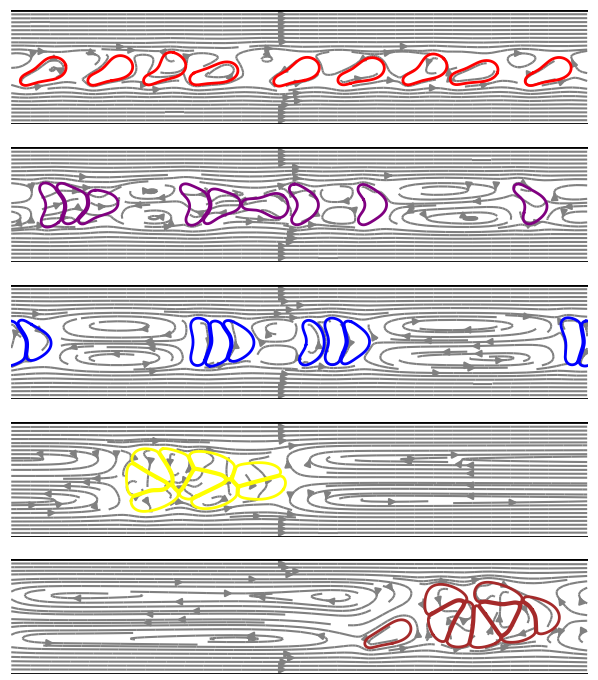}
\caption{\label{streamlines} Streamlines (grey lines with arrows) in a comoving frame. Snapshots showing the spatial configurations of the cells for different macroscopic adhesion energy values. The colors show different macroscopic adhesion energy values. Red : $0.0$, purple : $27.5$, blue : $53.0$ ,yellow : $105.0$ and brown : $158.0$. Here the viscosity contrast is set to $\lambda = 1.0$, $\phi = 0.1 \%$ and $C_{a} = 9.0$}
\end{figure}

Let us now provide an intuitive explanation for the non monotonic behavior of  of maximum flux with adhesion. At low adhesion some aggregates form, and they become larger and larger as adhesion increases. Figures \ref{fig5:3} and \ref{fig5:4} show typical configuration of RBCs for different hematocrits and adhesion energy. Consider for example the configurations in \ref{fig5:4} on the left. In the absence of adhesion (red) the RBCs form a file. Between cells (see Fig .\ref{streamlines}) we have flow recirculations.  Increasing adhesion (purple and blue) the cells start to form aggregates while remaining in the center. The cell-cell adhesion suppresses some recirculations (and thus dissipation), increasing thus efficiency of RBCs transport. When the adhesion energy increases further (yellow and brown) the aggregates become large enough so that they expend laterally to oppose enough resistance against the imposed flow. This leads to a decline in RBCs transport efficiency.
 

The fact that the optimal adhesion energy depends on capillary number has the following explanation. 
 When $Ca$ increases some aggregates are broken, and since there is a need to maintain some aggregate to destroy recirculation zones (and thus to enhance the flux) the adhesion energy has to be adapted (increased).

\subsection{Concentration profile}

\begin{figure}[hbtp]
\centering
\includegraphics[scale=0.55]{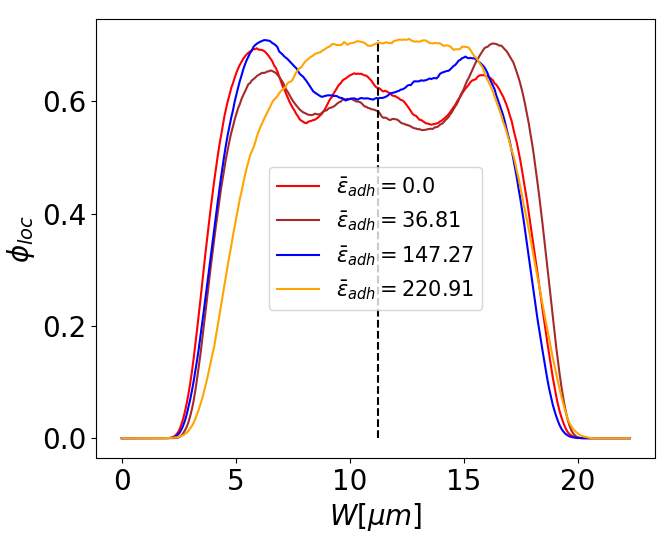}
\caption{\label{fig5:loc} Concentration profiles of suspension for different dimensionless adhesion energy values $\bar{\epsilon}_{adh}$. The black dashed line represents the channel center. Here $C_{a} = 25.0$ and $\phi_{t} = 46.0 \%$}
\end{figure}

\begin{figure}[hbtp]
\centering
\includegraphics[scale=0.16]{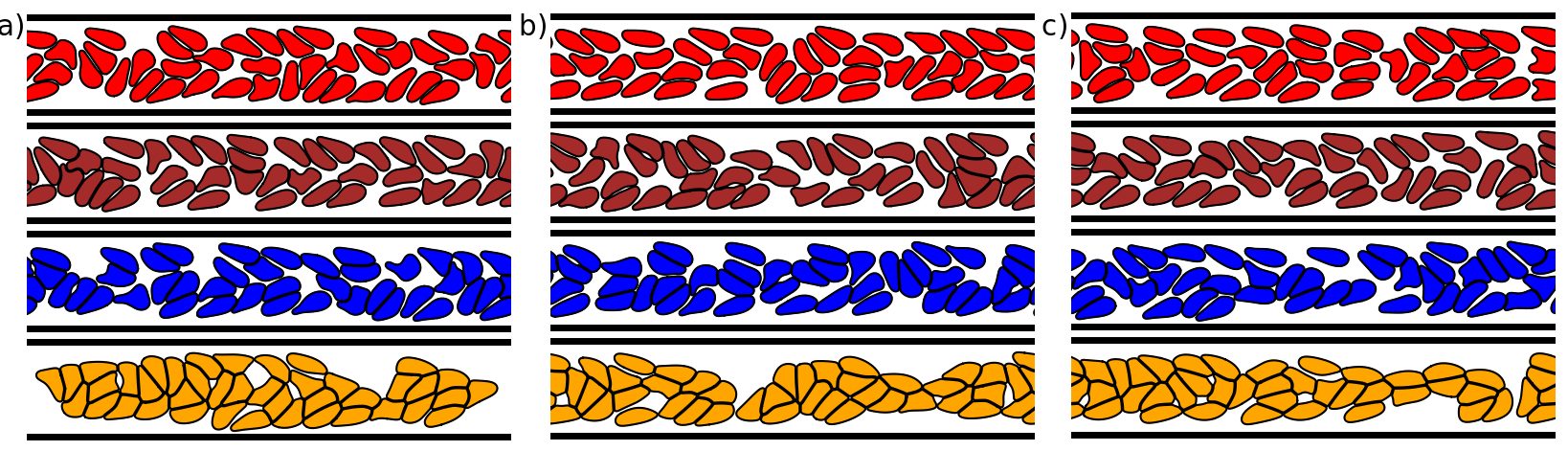}
\caption{\label{fig5:snap_41}  Snapshots showing the spatial configurations of the cells at different time steps. a) $1500 \tau_{c}$, b) $1750 \tau_{c}$ and c) $2000 \tau_{c}$. Here the capillary number is set to $C_{a} = 25.0$ , the cells concentration is $\phi_{t} = 46.0 \%$ and the viscosity contrast is set to $\lambda = 1.0$. The colors show different macroscopic adhesion energy values. Red : $0.0$, brown : $36.81$, blue : $147.27$ and orange : $220.91$}
\end{figure}

\begin{figure}[hbtp]
\centering
\begin{subfigure}
\centering
\includegraphics[scale=0.25]{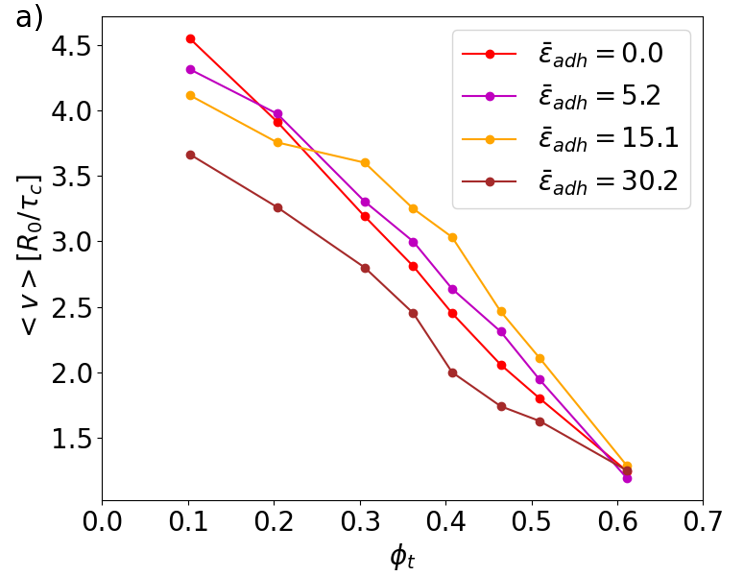}
\end{subfigure}
\hspace{0.1cm}
\begin{subfigure}
\centering
\includegraphics[scale=0.25]{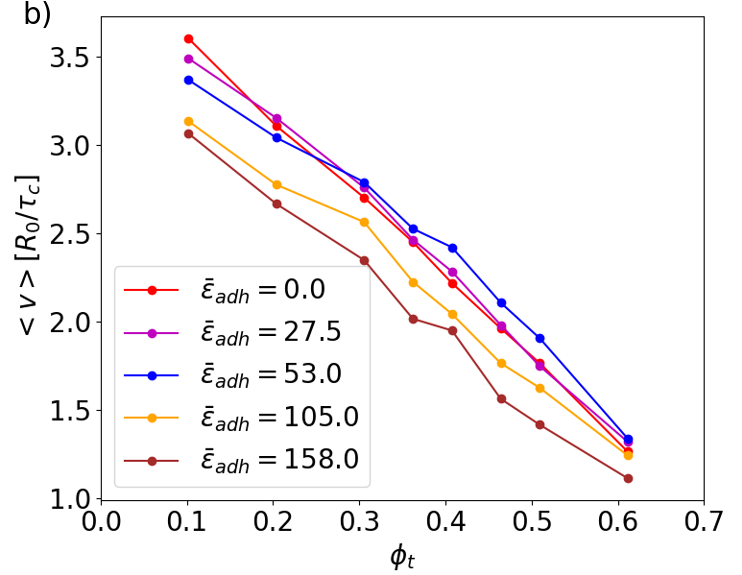}
\end{subfigure}

\begin{subfigure}
\centering
\includegraphics[scale=0.25]{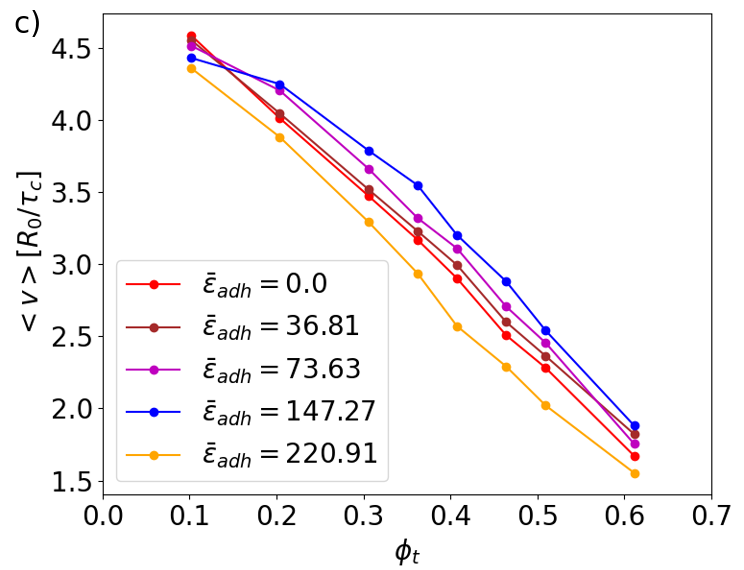}
\end{subfigure}
\hspace{0.1cm}
\begin{subfigure}
\centering
\includegraphics[scale=0.25]{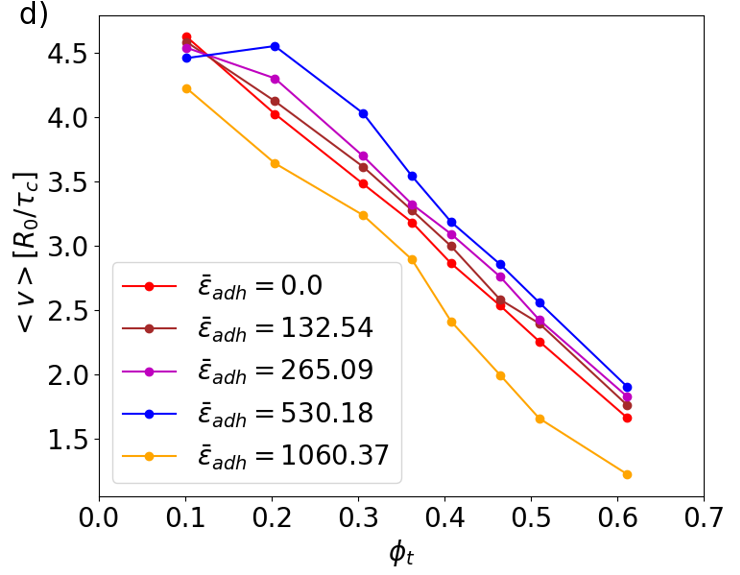}
\end{subfigure}
\caption{\label{fig5:5} The average velocity of vesicle mass center over cells and time as a function of volume fraction for channels with different dimensionless macroscopic adhesion energy. The simulation data are shown as dots. (a) $C_{a} = 0.9$, (b) $C_{a} = 9.0$, (c) $C_{a} = 25.0$, (d) $C_{a} = 90.0$.}
\end{figure}

Figure. \ref{fig5:loc} shows the concentration profiles of the suspension for different dimensionless adhesion energy values $\bar{\epsilon}_{adh}$. We see that in the absence of adhesion forces (red line in Figure. \ref{fig5:loc}), the concentration profile of the suspension exhibits three pics. This is due to the competition between two antagonist effects: i) cell-cell hydrodynamic interactions, leading to what is known as shear-induced diffusion,  pushing cells toward the walls, and ii) the wall-induced lift that pushes the cells towards the channel center. Increasing the adhesion energy in the physiological range (brown line in Figure. \ref{fig5:loc}), the concentration profile is not affected considerably. In the physiological range of adhesion energy the hydrodynamic interaction between cells prevails since the aggregates form in a reversible way. At high enough adhesion energy  (pathological case), large enough aggregates form (though they are still quite reversible, see snapshots in Fig.\ref{fig5:snap_41} ). In this case the concentration profile is shown on Fig. \ref{fig5:loc} (blue line). Comparing the red line (without adhesion) and the blue one in Fig. \ref{fig5:loc} (with large enough adhesion) one sees that around centerline (say between $y=5\mu $m and $y=15\mu $m the integrated concentration is larger for the blue one than for the red one. In addition, from Fig. \ref{fig5:5}c (same parameters as in Fig. \ref{fig5:loc}) the average speed in the channel are bigger for the blue line (with large enough adhesion; this increase of speed is due to dissipation reduction)  than for the red line (no adhesion). Together the increase of number of cells in the center and the average speed explain the increase of flux of RBCs. Increasing further and further adhesion leads to formation of large and irreversible aggregates occupying a larger and larger cross section, causing a relative blockage of the flow. The concentration profile (yellow line in Fig. \ref{fig5:loc}) has lost peaks where the aggregate accumulate in thee center due to the collapse of hydrodynamics diffusion. At the same time the average speed is reduced (Fig. \ref{fig5:1}a, yellow line), leading to a decline of the RBCs flux (Fig. \ref{fig5:1}a : brown curve, Fig. \ref{fig5:1}b : yellow and brown curves, \ref{fig5:1}c and d : yellow curve).

\subsection{The rheology of RBC aggregate suspension}

The rheology of human blood has been studied since decades\cite{faahraeus1929suspension, merrill1963rheology, chien1975biophysical, menu2000vivo, armstrong2004hydrodynamic, brust2014plasma}.  The most prominent feature is the shear-thinning behavior, associated with destruction of RBCs rouleaux under shear flow. We will see below some novel features associated to microcirculation of blood rheology, and especially the behavior of viscosity as a function of shear rate and hematocrit.

\begin{figure}[hbtp]
\centering
\begin{subfigure}
\centering
\includegraphics[scale=0.3]{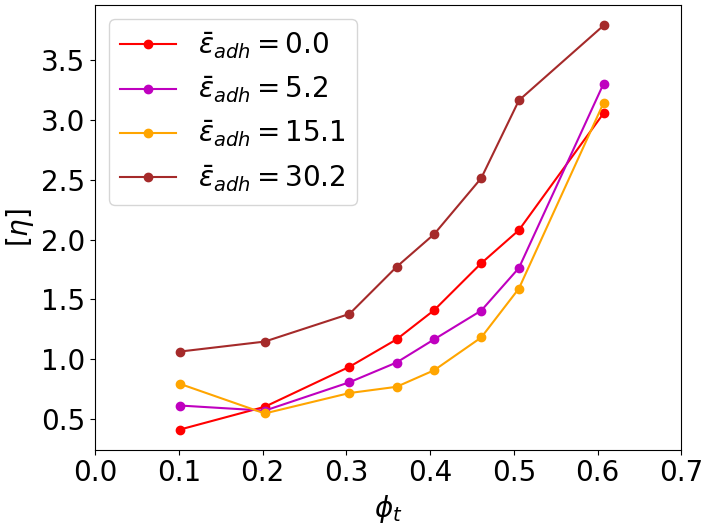}
\end{subfigure}
\hspace{0.1cm}
\begin{subfigure}
\centering
\includegraphics[scale=0.3]{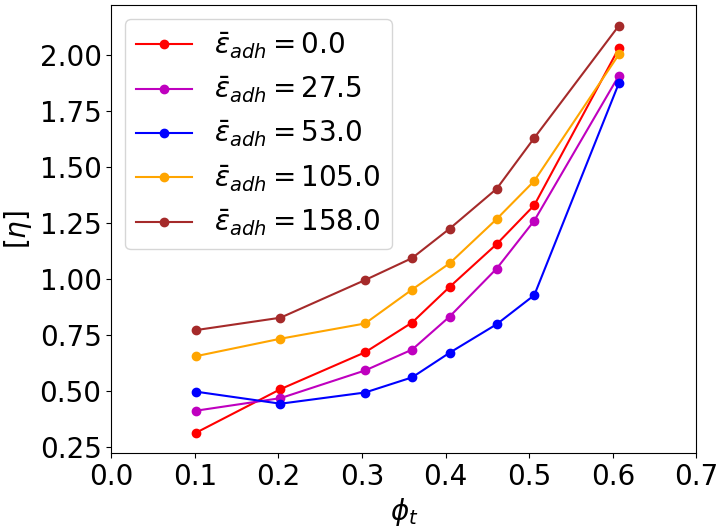}
\end{subfigure}

\begin{subfigure}
\centering
\includegraphics[scale=0.3]{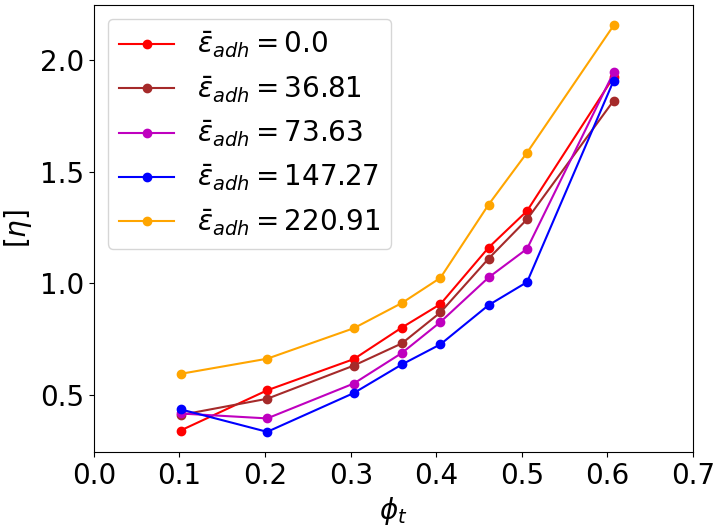}
\end{subfigure}
\hspace{0.1cm}
\begin{subfigure}
\centering
\includegraphics[scale=0.3]{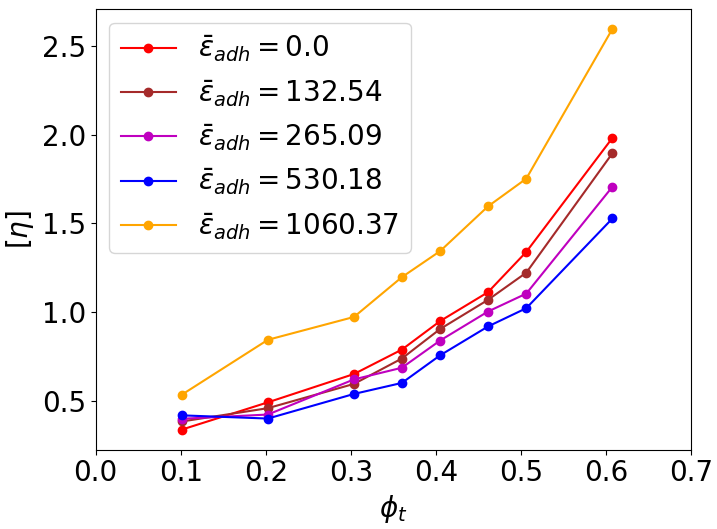}
\end{subfigure}
\caption{\label{fig5:6} The normalized viscosity as a function of volume fraction for channels with different dimensionless macroscopic adhesion energy. The simulation data are shown as dots. (a) $C_{a} = 0.9$, (b) $C_{a} = 9.0$, (c) $C_{a} = 25.0$, (d) $C_{a} = 90.0$. The reason of the presence of high values of adhesion energy ($\bar{\epsilon}_{adh} = 530.18, 1060.37$) is discussed in section \ref{sec1} }
\end{figure}

Let us first examine the behavior of the normalized viscosity (defined in \ref{poiseff}) as a function of hematocrit, for different adhesion energies.Figure. \ref{fig5:6} shows the results. We see clearly that that in a quite wide range of hematocrit the normalized viscosity decreases by increasing adhesion energy, before it increases again. This is in agreement regarding the analysis of RBCs flux. The difference between the minimal viscosity (the lower curves in Figure. \ref{fig5:6}) and the viscosity without adhesion (red lines in Figure. \ref{fig5:6}) may attain about 50$\%$. We believe that the origin of this behavior is similar to that given for the non monotonic behavior of RBCs flux with the adhesion energy.  In Fig. \ref{eta_adh} we show (for a given hematocrit) the behavior of the normalized viscosity as a function of the adhesion energy. This figure exhibits the existence of a specific value of adhesion for which the viscosity is minimal. This specific adhesion depends on the capillary number. 

{Blood is a shear-thinning fluid, meaning that the blood viscosity decreases with shear rate. This is due to dissociation of RBCs upon increasing shear rate. In the presence of adhesion between cells
 the shear-thinning is observed  (see in Fig .\ref{fig:eta}). When $\lambda = 1.0$, the aggregates dissociate at high shear rate. This implies (see in Fig .\ref{fig:eta} (a)), all the curves, corresponding to different adhesion energies,  collapse on each other at high shear rate. Contrary to the case $\lambda = 1.0$, at high enough viscosity contrast value $\lambda = 10.0$, the aggregates are quite robust. For this reason the curves in Fig .\ref{fig:eta} (b) do not collapse on each other at high shear rate.

\begin{figure}[hbtp]
\centering
\includegraphics[scale=0.55]{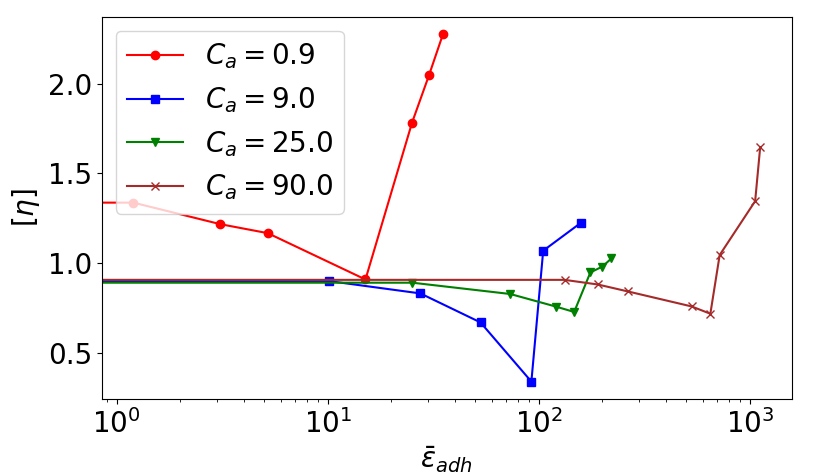}
\caption{\label{eta_adh} The normalized viscosity as function of the dimensionless macroscopic adhesion energy for different values of capillary number. The horizontal axis is in logarithmic scale. here the cell concentration is $\phi_{t} = 46.0 \%$.}
\end{figure}

\begin{figure}[hbtp]
\centering
\includegraphics[scale=0.55]{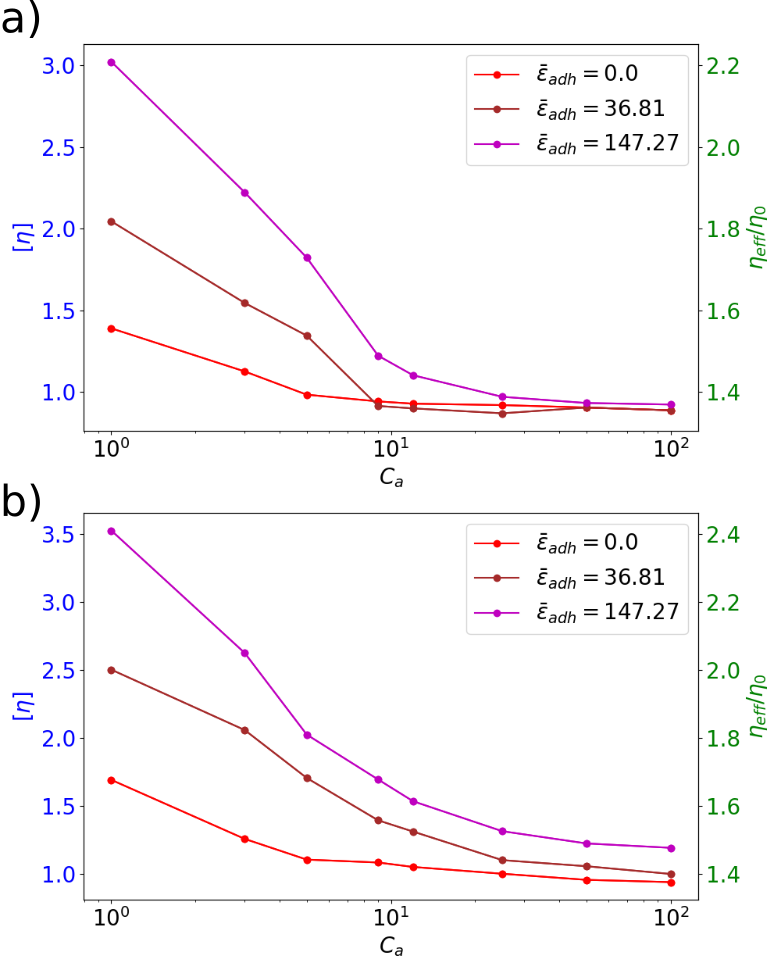}
\caption{\label{fig:eta}
 The normalized viscosity $[\eta]$ (the blue axis) and the effective viscosity $\eta_{eff}$ (green axis) as a function of capillary number with different dimensionless macroscopic adhesion energy. The simulation data are shown as dots. The cell concentration is $\phi_{t}=40\%$. (a) $\lambda = 1.0$, (b) $\lambda = 10.0$ 
}
\end{figure}

\section{Conclusions}
A major finding of this article is the existence of an optimal adhesion energy corresponding to a maximum flux  of cells. We have provided a qualitative argument to explain this counter-intuitive behavior. It is a combination of recirculation zones (see Fig. \ref{streamlines}) that are suppressed due to aggregate formation (this lowers dissipation) and the fact that large aggregates (due to a high enough adhesion)  tend to block the channel, and cause an obstacle for cell flow.  The interplay between these two effects leads to an optimal adhesion energy. The non monotonic behavior  is also exhibited by the effective viscosity of the suspension. We have also seen that adhesion can boost transport of RBCs to a level attaining up to about 30 $\%$ in comparison to the case without adhesion. In vivo, it is known that aggregates can form and can be broken by hydrodynamic stress, except in pathological situations where aggregates become robust and may tend to cause vessel occlusion. We have learnt from this study that an adhesion, which is not too strong, may be beneficial for perfusion. 
 
 This work has several limitations. Firstly, it was performed  for a 2D study where some real physics of RBC is absent, such as cytoskeletton. This study could be viewed as a guide for a systematic 3D exploration. 
 The fact that the phase diagram obtained in \cite{abbasi2021erythrocyte} (regarding doublet dynamics) for 3D is strikingly similar to that obtained in 2D lends confidence that the 2D study analyzed in this article could, at least, have a qualitative value. Secondly, the study focused on a straight geometry, which is far away from real vascular networks. It will be imperative to analyze the various spatio-temporal properties, as well as transport phenomena occurring in real vascular networks. It is hoped that what we have learnt so far on a simplistic geometry and in 2D, will serve to identify relevant features to be studied in complex geometries. 


\section*{Conflicts of interest}
There are no conflicts to declare.

\section*{Acknowledgements}
We thank CNES (Centre National d’Etudes Spatiales) for financial support and for having access to data of microgravity, and the French-German university programme “Living Fluids” (Grant CFDA-Q1-14) for financial support. 





\end{document}